\documentclass{appolb}
\usepackage{eurosym}
\usepackage{graphicx}
\usepackage{amsmath}
\usepackage{multirow}
\newcommand {\nc} {\newcommand}
\nc {\beq} {\begin{eqnarray}} \nc {\eol} {\nonumber \\} \nc {\eeq}
{\end{eqnarray}} \nc {\eeqn} [1] {\label{#1} \end{eqnarray}} \nc
{\eoln} [1] {\label{#1} \\} \nc {\ve} [1] {\mbox{\boldmath $#1$}}
\nc {\rref} [1] {(\ref{#1})} \nc {\Eq} [1] {Eq.~(\ref{#1})} \nc
{\re} [1] {Ref.~\cite{#1}} \nc {\dem} {\mbox{$\frac{1}{2}$}} \nc
{\arrow} [2] {\mbox{$\mathop{\rightarrow}\limits_{#1 \rightarrow
#2}$}}

\begin{document}
\title{ASTROPHYSICAL S(0)-FACTORS FOR THE $^{3}{\rm He}(\alpha,
\gamma)^{7}{\rm Be}$, $^{3}{\rm H}(\alpha, \gamma)^{7}{\rm Li}$ and
$^{7}{\rm Be}(p, \gamma)^{8}{\rm B}$ DIRECT CAPTURE PROCESSES IN A
POTENTIAL MODEL
\thanks{%
Presented at the IV International Scientific Forum "Nuclear Science
and Technologies", Almaty, Kazakhstan, 26-30 September, 2022. }}
\author{S.A.Turakulov, E.M.Tursunov
\address{{Institute of Nuclear Physics, 100214, Tashkent, Uzbekistan}     \\
 {National University of Uzbekistan, 100174, Tashkent, Uzbekistan}}}
\maketitle
\begin{abstract}
Astrophysical S-factors at zero energy for the direct nuclear
capture reactions $^{3}{\rm He}(\alpha, \gamma)^{7}{\rm Be}$,
$^{3}{\rm H}(\alpha, \gamma)^{7}{\rm Li}$ and $^{7}{\rm Be}(p,
\gamma)^{8}{\rm B}$ are estimated within the framework of two-body
potential cluster model on the basis of extranuclear capture
approximation of D. Baye and E. Brainis. The values of S(0)-factors
have been calculated using two different potential models for each
process, which were adjusted to the binding energies and empirical
values of the asymptotical normalization coefficients from the
literature. New values of S(0)-factors have been obtained.
\end{abstract}
%
%

\section{Introduction}
Determination of the low-energy values of the astrophysical S-factor
for the direct radiative capture reactions $d(\alpha,
\gamma)^{6}{\rm Li}$, $^{3}{\rm He}(\alpha, \gamma)^{7}{\rm Be}$,
$^{3}{\rm H}(\alpha, \gamma)^{7}{\rm Li}$ and $^{7}{\rm Be}(p,
\gamma)^{8}{\rm B}$, especially at E=0, plays an important role in
nuclear astrophysics in the both Standard Solar and Big Bang
nucleosynthesis (BBN) models \cite{adelber11,fields11}. The
calculation of S(0) is carried out only with the help of theoretical
approaches, since the direct experimental measurements of the
cross-section at ultralow energies are not possible due to very
small values of the cross-section. In particular, for the first
capture reaction at energies of 10 keV, the cross-section is of the
order of nanobarns. As it is well-known, the $\alpha+d\rightarrow
^6$Li$+\gamma$ synthesis process is the main source of the $^6$Li
isotope in a period of primordial nucleosynthesis.
 For this reason, investigating the $\alpha+d\rightarrow ^6$Li$+\gamma$
synthesis is of a great interest within the experimental
\cite{robert81,kiener91,mohr94,luna14,luna17} and theoretical
studies \cite{dub10,bt18,sol22}. The direct capture processes
$^{3}{\rm He}(\alpha, \gamma)^{7}{\rm Be}$ and $^{3}{\rm H}(\alpha,
\gamma)^{7}{\rm Li}$  present the main source of the primordial
$^7$Li element \cite{tur18}. All the above mentioned reactions are
directly related to the cosmological lithium problem\cite{asp06}.
Moreover, the processes $^{3}{\rm He}(\alpha, \gamma)^{7}{\rm Be}$
and $^{7}{\rm Be}(p, \gamma)^{8}{\rm B}$ are essential for the
estimation of neutrino fluxes from the Sun
\cite{adelber11,fields11}. In the present work, we extend the
theoretical model previously developed in Refs.
\cite{tur18,tur15,tur20,tur21,tur21p} for the determination of zero
energy astrophysical S-factor on the basis of extranuclear capture
approximation as proposed in Refs.\cite{baye98,baye20}. The aim of
present work is to determine the values of the S(0)-factor for the
 $^{3}{\rm He}(\alpha,
\gamma)^{7}{\rm Be}$, $^{3}{\rm H}(\alpha, \gamma)^{7}{\rm Li}$ and
$^{7}{\rm Be}(p, \gamma)^{8}{\rm B}$ direct nuclear capture
reactions within the framework of two-body potential cluster model.
The method can be  extended to the direct capture process $d(\alpha,
\gamma)^{6}{\rm Li}$ within the "exact-mass prescription". However
the last case presents only the methodical interest, since the
isospin forbidden E1 astrophysical  S factor of the process
$d(\alpha, \gamma)^{6}{\rm Li}$ can be described only within the
three-body model, but not in the two-body model \cite{bt18}.

\section{Theoretical model}

In fact, the experimental measurements and theoretical approaches
define the cross-sections of the process. But, in the capture process which involves light nuclei,
the cross-section decreases exponentially when the energy tends to zero. Therefore, in the low energy nuclear astrophysics
the astrophysical $S$-factors are used. This
quantity is expressed with the help of the cross section as
\cite{tur15,nacre99}
\begin{eqnarray}
S(E)=\sum_{l_f J_f}S_{l_f J_f}(E)= E \,\exp(2 \pi \eta)\sum_{l_f
J_f}\sum_{l_i J_i} \sum_{\lambda} \sigma^{\Omega\lambda}_{l_i J_i
\rightarrow l_f J_f}(E) \label{eq1},
\end{eqnarray}
where $l_i,J_i$ ($l_f,J_f$) are the orbital and the total angular
momenta of the initial (final) states, respectively, $\eta$ is
the Zommerfeld parameter, $\Omega=$ E or M (electric or magnetic
transition), $\lambda$ is a multiplicity of the transition. For the
above radiative capture reactions the electric dipole E1 and
quadrupole E2 transitions contributions are dominant in the
cross-section. Thus, the cross sections for the electric transitions
of the radiative capture process is expressed as
\cite{baye98,nacre99}
\begin{eqnarray}
\sigma^{E\lambda}_{l_i J_i \rightarrow l_f J_f}(E)= \frac{8 \pi e^2
\mu}{\hbar c}\frac{k_{\gamma}^{2 \lambda+1}}{ k^3} \cdot N_{E
\lambda} \cdot \left[I_{if}(E)\right]^2 \label{eq2}
\end{eqnarray}
where $k=\sqrt{2\mu E}/\hbar c$ is the wave number of the colliding
particles relative motion, $\mu$ is the reduced mass of the clusters
involved in the capture process. $k_{\gamma}=E_\gamma / \hbar c$ is
the wave number of the photon corresponding to energy
$E_\gamma=E_{\rm th}+E$, where $E_{\rm th}$ is the threshold energy
and $N_{E \lambda}$ is
\begin{eqnarray}
N_{E \lambda}=\left[Z_1 \left( \frac{A_2}{A} \right)^{\lambda}+Z_2
\left(\frac{-A_1}{A} \right)^{\lambda} \right]^2
\frac{\lambda(\lambda+1)[\lambda][l_i][J_i][J_f]} {\left( \left[
\lambda \right]!! \right)^2 [S_1][S_2]}
\\
\nonumber \times \left(C^{l_f 0}_{\lambda 0 l_i 0}
\right)^{2}\left\{
\begin{array}{ccc}
J_i & l_i & S \\
l_{f} & J_{f} & \lambda
\end{array} \right\}^2.  \label{eq3}
\end{eqnarray}
The parameters $A_1$, $A_2$  are mass numbers of the clusters in the entrance
channel, $A=A_1+A_2$, $S_1$, $S_2$ are spins of the clusters, $S$ is
a spin of reaction channel. We also use short-hand notations
$[S]=(2S+1)$and $[\lambda]!!=(2\lambda+1)!!$. The overlap integral
is given as
\begin{eqnarray}
I_{if}(E)=\int^{\infty}_{0} u_{E}^{(l_f S J_f)}(r)r^{\lambda}u^{(l_i
S J_i)}(E,r) dr.\label{eq4}
\end{eqnarray}
where $u_{E}^{(l_f S J_f)}(r)$ and $u^{(l_i S J_i)}(E,r)$ are final
bound and initial scattering wave functions, respectively.
\par At next step, we determine the zero energy astrophysical
S(0)-factor. In order to distinguish energy dependent parts of the
astrophysical $S$-factor we introduce modified  scattering functions \cite{baye98}
\begin{eqnarray}
\tilde{u}^{(l_i S J_i)} (E,r)=E^{1/2}\exp{(\pi\eta)}u^{(l_i S J_i)}
(E,r), \label{eq5}
\end{eqnarray}
here $E^{1/2}=\frac{k \cdot \hbar c}{\sqrt{2\mu}}$ is kinetic energy
of the relative motion. When $E$ tends to zero, consequently $\eta$
also tends to infinity and the regular $F_l$ and irregular $G_l$
Coulomb wave functions become unusable. Therefore, the radial
scattering wave function is normalized with the help of the rescaled Coulomb functions
$\mathcal{F}_l$ and $\mathcal{G}_l$ \cite{baye20} as
\begin{eqnarray}
\tilde{u}^{(l_i S J_i)}(E,r) \arrow{r}{\infty} \cos\delta_{(l_i S
J_i)}(E) \mathcal{F}_l(E,r) +
\frac{2}{\pi}\exp(2\pi\eta)\sin\delta_{(l_i S J_i)}(E)
\mathcal{G}_l(E,r), \label{eq6}
\end{eqnarray}
where, $\delta_{l_i S J_i}(E)$ is the phase shift in the $(l,S,J)$th
partial wave. This normalization provides that $\tilde{u}^{(l_i S
J_i)}(E,r)$ has a finite limit when $E$ tends to zero
\cite{baye98,baye20}. It will be convenient to make use of a
function of the phase shift $\delta_{l_i S J_i}(E)$ defined as
\begin{eqnarray}
\mathcal{D}_{l_i S J_i}
(E)=\frac{2}{\pi}\left[\exp(2\pi\eta)-1\right]\tan \delta_{l_i S
J_i} (E) \label{eq7}
\end{eqnarray}
which also has a finite limit when $E\rightarrow 0 $ \cite{baye20}.
Taking into accounted that at the ultralow energy the phase shift
$\delta_{l_i S J_i}(E)$ is very small and satisfies the condition
$\exp(-2\pi\eta)<<1$, the asymptotic form (\ref{eq6}) of the radial
wave function becomes
\begin{eqnarray}
\tilde{u}^{(l_i S J_i)}(E,r) \arrow{r}{\infty} \mathcal{F}_l(E,r) +
\mathcal{D}_{l_i S J_i}(E) \mathcal{G}_l(E,r), \label{eq8}
\end{eqnarray}
which remains finite at $E=0$.
\par Finally, we can rewrite the expression for the zero energy astrophysical S(0)-factor in the form \cite{baye98,baye20}
\begin{eqnarray}
S(0)=\frac{1}{2}\alpha \hbar c \left( \frac{E_{th}}{\hbar
c}\right)^{2 \lambda+1}\cdot N_{E \lambda} \cdot
\left[I_{if}(0)\right]^2 \label{eq9}.
\end{eqnarray}
where $\alpha$ is the fine-structure constant. The zero energy
overlap integral is given as
\begin{eqnarray}
I_{if}(0)=\int^{\infty}_{0} u_{E}^{(l_f S J_f)}(r)r^{\lambda}
\tilde{u}^{(l_i S J_i)}(0,r) dr.\label{eq10}
\end{eqnarray}
where,
\begin{eqnarray}
\tilde{u}^{(l_i S J_i)}(0,r) \arrow{r}{\infty} \mathcal{F}_l^{0} (r)
+ \mathcal{D}_{l_i S J_i}(0) \mathcal{G}_l^{0} (r), \label{eq11}
\end{eqnarray}
Using above equations one is able to estimate
the astrophysical S(0)-factor of above
capture processes within the two-body cluster model.
\section{Results and discussion}
Calculations of the cross section and astrophysical S(0)-factor have
been performed under the same conditions as in
Refs.\cite{tur18,tur15,tur20,tur21,tur21p}. The Schr\"{o}dinger
equation in the entrance and exit channels is solved with the
two-body central nuclear potentials of the Gaussian form
\cite{dub10} with the corresponding point-like Coulomb potential for
the $\alpha+d$ and $p+ ^{7}{\rm Be}$ systems as \cite{dub10,dub19}.
For synthesis of  $^{3}{\rm He}+\alpha$ and $^{3}{\rm H}+\alpha$
have been used spherical form of Coulomb potential. For consistency
we use the same model parameters as in the aforementioned paper
\cite{tur15}: $\hbar^2/2$[a.m.u]=20.7343 MeV fm$^2$. The  Coulomb
parameter ${\rm R}_c$ = 3.095 fm for the spherical form of potential
\cite{tur18}. The mass number corresponding for the first $A_1$
particle is $m_{\rm d}=$2.0 a.m.u., $m_{^{3}{\rm He}}=m_{^{3}{\rm
H}}=$3.0 a.m.u., $m_{p}=$1.007 276 4669 a.m.u. and the mass number
$A_2$ of the second particle is $m_\alpha=$ 4.0 a.m.u., $m_{^7{\rm
Be}}=$ 7.014735 a.m.u., respectively.

It should be noted, that for the calculations of the $\alpha+d$
capture reaction we use the "exact mass" prescription in the
two-body model. As was noted in the Introduction, this case is of
the methodical interest, since realistic estimates of the
isospin-forbidden E1 S-factor can be obtained only within the
three-body model \cite{bt18,tur18,tur20,tur21}. Thus, within the
assumption of the "exact mass" prescription the exact experimental
mass values $m_{\rm d}=A_1=$2.013553212724 a.m.u., $m_\alpha=A_2=$
4.001506179127 a.m.u. \cite{dub10}  are used.

\begin{table*}[htbp]
\caption{ The values of ANC for the $\alpha+d \to ^{6}{\rm Li}$
virtual transition and corresponding astrophysical S(0)-factors of
the direct $d(\alpha, \gamma)^{6}{\rm Li}$ capture reaction.}
\label{tab1}
\begin{center}
\begin{tabular}{|c|c|c|} \hline
Model & $C_{\alpha d}$, fm$^{-1/2}$ & S(0), MeV nb\\ \hline
${\rm V}_{\rm D}$ & 2.53 & 1.53 \\
\hline ${\rm V}_{\rm M}$  & 2.31 & 1.26 \\ \hline
\end{tabular}
\end{center}
\end{table*}

\par The obtained S(0)-factor values for the above capture processes
are strongly dependent on the value of the asymptotical normalization coefficient
(ANC). The values of ANC of the $\alpha+d \to ^{6}{\rm Li}$ virtial transition and
astrophysical S(0)-factors of the  direct $d(\alpha, \gamma)^{6}{\rm
Li}$ capture process are presented in Table \ref{tab1} for the two potential models ${\rm
V}_{\rm D}$ \cite{dub10} and ${\rm V}_{\rm M}$ \cite{tur15}. The initial potential ${\rm
V}_{\rm D}$   yields a value $C_{\alpha d}=2.53$
fm$^{-1/2}$ for the ANC. The modified potential ${\rm V}_{\rm M}$
 yields $C_{\alpha d}=2.31$ fm$^{-1/2}$, which is more
consistent with the empirical value of ANC, $C_{\alpha
d}=2.32\pm0.11$ fm$^{-1/2}$ extracted from the experimental data in
Ref. \cite{blok93}.

\begin{table*}[htbp]
\caption{Values of ANC for the ground $p_{3/2}$ and first excited
$p_{1/2}$ bound states of the $^{7}{\rm Be}$ and $^{7}{\rm Li}$
nuclei and corresponding astrophysical S(0)-factors.} \label{tab2}
\begin{center}
\begin{tabular}{|c|c|c|c|c|} \hline
Reaction & Model & $C_{p_{3/2}}$, fm$^{-1/2}$ & $C_{p_{1/2}}$,
fm$^{-1/2}$ &  S(0), keV b\\ \hline
\multirow{2}{*}{$^{3}{\rm
He}(\alpha, \gamma)^{7}{\rm Be}$} & ${\rm
V}_{\rm D}^{n}$ & 4.34 & 3.71 & 0.56 \\
& ${\rm V}_{\rm M1}^{n}$ & 4.79 & 4.24 & 0.58 \\ \hline
\multirow{2}{*}{$^{3}{\rm
H}(\alpha, \gamma)^{7}{\rm Li}$} & ${\rm V}_{\rm D}^{n}$ & 3.72 & 3.12 & 0.10\\
& ${\rm V}_{\rm M1}^{n}$ & 4.10 & 3.55 & 0.09\\ \hline
\end{tabular}
\end{center}
\end{table*}

In Table \ref{tab2} the values of ANC for the ground $p_{3/2}$ and the first excited
$p_{1/2}$ bound states of the $^{7}{\rm Be}(=^{3}{\rm He}+\alpha)$ and $^{7}{\rm Li}(=^{3}{\rm
H}+\alpha)$ nuclei and calculated results for corresponding astrophysical S(0)-factors
are given within two potential models ${\rm V}_{\rm D}^{n}$ and ${\rm V}_{\rm M1}^{n}$  \cite{tur18, tur21}.
The parameters of these potential models are given in Ref.\cite{tur21}.

These models differ each from other only with values of ANC for the
bound states of $^7$Be and $^7$Li nuclei. They have been adjusted to
the values of ANC extracted from the analysis of the low energy
experimental astrophysical S-factors of the $^{3}{\rm He}(\alpha,
\gamma)^{7}{\rm Be}$ and $^{3}{\rm H}(\alpha, \gamma)^{7}{\rm Li}$
reactions in Refs. \cite{qah12,igam07}. The obtained theoretical
estimations of the zero energy S(0)-factor for the $^{3}{\rm
He}(\alpha, \gamma)^{7}{\rm Be}$ process with the potentials ${\rm
V}_{\rm D}^{n}$ and ${\rm V}_{\rm M1}^{n}$ are very consistent with
the new data S(0)=0.56$\pm$0.04 keV b of the Solar Fusion II
Collaboration \cite{adelber11}. In addition, obtained results for
the $^{3}{\rm H}(\alpha, \gamma)^{7}{\rm Li}$ capture process in the
frame of proposed couple of potential models are in a good agreement
with the result S(0)=0.10 $\pm$0.02 keV b of the NACRE
Collaboration\cite{nacre99}.

\begin{table*}[htbp]
\caption{ The values of ANC for the $p+ ^{7}{\rm Be}\to ^{8}{\rm B}$
virtual transition and corresponding astrophysical S(0)-factors of
the direct $^{7}{\rm Be}(p, \gamma)^{8}{\rm B}$ capture reaction.}
\label{tab3}
\begin{center}
\begin{tabular}{|c|c|c|} \hline
Model & $C_{\alpha d}$, fm$^{-1/2}$ & S(0), eV b\\ \hline
${\rm V}_{\rm D}$ & 0.70 & 18.32 \\
\hline ${\rm V}_{\rm M}$  & 0.73 & 19.61 \\ \hline
\end{tabular}
\end{center}
\end{table*}
In Table \ref{tab3} the values of ANC for the bound state of the
$^{8}{\rm B}(=p+ ^{7}{\rm Be})$ nucleus and calculated results for
corresponding astrophysical S(0)-factors are given for two potential
models ${\rm V}_{\rm D}$ and ${\rm V}_{\rm M}$ from
Ref.\cite{tur21p}, where the full set of parameters of these
potential models are given. As can be seen from the table, the
obtained theoretical estimates for the zero energy S(0)-factor of
the $^{7}{\rm Be}(p, \gamma)^{8}{\rm B}$ process are consistent with
the new data S(0)=20.8$\pm$2.10 eV b of the Solar Fusion II
Collaboration \cite{adelber11}.

\section{Conclusions}

In the present work, the zero energy astrophysical S-factors for the
 $^{3}{\rm He}(\alpha, \gamma)^{7}{\rm Be}$, $^{3}{\rm H}(\alpha,
\gamma)^{7}{\rm Li}$ and $^{7}{\rm Be}(p, \gamma)^{8}{\rm B}$
radiative capture reactions have been estimated in the framework of
the two-body potential cluster model on the basis of extranuclear
capture approximation \cite{baye98,baye20}. For the astrophysical
S(0)-factors of the $^{3}{\rm He}(\alpha, \gamma)^{7}{\rm Be}$,
$^{3}{\rm H}(\alpha, \gamma)^{7}{\rm Li}$ and $^{7}{\rm Be}(p,
\gamma)^{8}{\rm B}$ capture reactions the obtained estimates are
consistent with new data sets of the Solar Fusion II and NACRE
Collaborations.

\section{Acknowledgements}
The authors acknowledge Prof. Daniel Baye for useful discussions and
valuable advice.


\begin{thebibliography}{99}
\bibitem{adelber11} E.G. Adelberger  et al.,  Rev. Mod. Phys. 83, 195 (2011).
\bibitem{fields11} B.D. Fields, Annual Review of Nuclear and Particle Science 61, 47
(2011).

\bibitem{robert81}
R.G.H. Robertson et al. Phys. Rev. Lett. 47, 18 (1981).
\bibitem{kiener91}
J.Kiener et al. Phys. Rev. C 44, 21 (1991).
\bibitem{mohr94} P. Mohr et
al. Phys. Rev. C 50, 15 (1994).
\bibitem{luna14} M. Anders et al. (LUNA Collaboration), Phys. Rev. Lett. 113,
042501 (2014).
\bibitem{luna17}
D. Trezzi et al. (LUNA Collaboration), Astroparticle Physics 89, 57
(2017).
\bibitem{dub10}
S.B.Dubovichenko, Phys. Atomic Nucl. 73, 1526 (2010).
\bibitem{bt18} D.Baye and E.M.Tursunov,  J. Phys. G: Nucl. Part. Phys.  45,  085102 (2018).
\bibitem{sol22} A.S.Solovyev, Phys. Rev. C 106, 014610 (2022).
\bibitem{tur18} E.M.Tursunov, S.A.Turakulov, A.S.Kadyrov, Phys. Rev. C 97, 035802 (2018).
\bibitem{asp06} M. Asplund, D. L. Lambert, P. E. Nissen, F. Primas, and V. V.
Smith, The Astrophysical Journal  644, 229 (2006).
\bibitem{tur15}
E.M.Tursunov, S.A.Turakulov, P.Descouvemont. Phys. Atom. Nucl. 78,
193 (2015).
\bibitem{tur20}
E.M.Tursunov, S.A.Turakulov, A.S.Kadyrov, Nuclear Physics A 1000,
121884 (2020).
\bibitem{tur21}
E.M.Tursunov, S.A.Turakulov, A.S.Kadyrov, Nuclear Physics A 1006,
122108 (2021).
\bibitem{tur21p} E.M.Tursunov, S.A.Turakulov, A.S.Kadyrov, L.D.Blokhintsev, Phys. Rev. C 104, 045806 (2021).
\bibitem{baye98} D.Baye, P.Descouvemont, M.Hesse,
Phys. Rev. C 58, 545 (1998).
\bibitem{baye20} D.Baye, E.Brainis,
Phys. Rev. C 61, 025801 (2000).
\bibitem{nacre99}
C. Angulo {\it et~al.} (NACRE),  Nuclear Physics A 656, 3 (1999).
\bibitem{dub19} S. B.Dubovichenko, N. A.Burkova, A. V.Dzhazairov-Kakhramanov, A. S.Tkachenko,
 Nucl. Phys. A 983, 175 (2019).
\bibitem{blok93} L.D.Blokhintsev, V.I.Kukulin, A.A.Sakharuk,
D.A.Savin, E.V.Kuznetsova, Phys. Rev.  C 48, 2390 (1993).
\bibitem{qah12}
Q.I. Tursunmahatov, R. Yarmukhamedov, Phys. Rev. C 85, 045807
(2012).
\bibitem{igam07} S.B. Igamov, R. Yarmukhamedov, Nucl. Phys. A 781, 247
(2007).
\end{thebibliography}

\end{document}